\documentstyle[12pt]{article}
\begin{document}
\begin{center}
{\Large
Aging in a simple glassformer\footnote{Talk presented at ``Unifying
Concepts in Glass Physics'', ICTP, Trieste 15 - 18 September 1999}
}

Walter Kob$^{+}$, Jean-Louis Barrat$^{\dag}$, Francesco 
Sciortino$^{o}$ and Piero Tartaglia$^{o}$
\end{center}

\noindent
$^{+}$ Institute of Physics, Johannes-Gutenberg University,
D-55099 Mainz, Germany\\
\noindent
$^{\dag}$D\'epartement de Physique des Mat\'eriaux,
Universit\'e Claude Bernard and CNRS, 69622 Villeurbanne Cedex, France\\
\noindent
$^{o}$ Dipartimento di Fisica and Istituto Nazionale
per la Fisica della Materia, Universit\'a di Roma {\it La Sapienza},
P.le Aldo Moro 2, I-00185 Roma, Italy.

\vspace*{5mm}
\par
{\bf Abstract}
Using molecular dynamics computer simulations we investigate the
out-of-equilibrium dynamics of a Lennard-Jones system after a quench
from a high temperature to one below the glass transition temperature. By
studying the radial distribution function we give evidence that during the
aging the system is very close to the critical surface of mode-coupling
theory. Furthermore we show that two-time correlation functions show
a strong dependence on the waiting time since the quench and that their
shape is very different from the one in equilibrium. By investigating
the temperature and time dependence of the frequency distribution of
the normal modes we show that the energy of the inherent structures
can be used to define an effective (time dependent) temperature of the
aging system.  \bigskip

\section{Introduction}
In the last few years ample evidence has been accumulated that the
mode-coupling theory of the glass transition (MCT) gives a reliable
description of the dynamics of simple supercooled liquids on a qualitative
as well as quantitative level~\cite{gotze99}. Recently it has even been
documented that also some aspects of the dynamics of {\it strong} glass
formers are described well by the theory~\cite{franosch97horbach99}. Thus
we can conclude that many of the key aspects of the dynamics of
supercooled liquids are understood in a quite satisfactory way. This
is not yet the case for the dynamics of glasses {\it below} the glass
transition temperature, i.e. in that temperature regime in which the {\it
equilibrium} relaxation time significantly exceeds the time scale of
the experiment. Only relatively recently first attempts have been made
to understand this out-of-equilibrium dynamics within the framework of
statistical mechanics and thermodynamics~\cite{aging_theo}. In particular
it was found that, for certain systems, the equations of motion describing
the dynamics below $T_g$ are formally quite similar to the MCT equations,
which, as discussed above, describe well the relaxation dynamics {\it
above} $T_g$. Whether or not these out-of-equilibrium theories will be
equally successful to describe the dynamics of structural glasses below
$T_g$ is currently not known and in the present paper we discuss some
computer simulations which have been done to test these theories.

\section{Model and Details of the Simulations}

The system we consider is a binary (80:20) mixture of particles
which interact with a Lennard-Jones potential, $V_{\alpha\beta}=
4\epsilon_{\alpha\beta}[(\sigma_{\alpha\beta}/r)^{12}-
(\sigma_{\alpha\beta}/r)^6]$. Here $\alpha,\beta\in \{\rm A,B\}$ denote
the species of the particles and the parameters $\epsilon_{\alpha\beta}$
and $\sigma_{\alpha\beta}$ are given by $\epsilon_{AA}=1.0$,
$\sigma_{AA}=1.0$, $\epsilon_{AB}=1.5$, $\sigma_{AB}=0.8$,
$\epsilon_{BB}=0.5$, and $\sigma_{BB}=0.88$. This potential is
truncated and shifted at a distance $\sigma_{\alpha\beta}$. In the
following we will use $\sigma_{\rm AA}$ and $\epsilon_{\rm AA}$ as the
unit of length and energy, respectively (setting the Boltzmann constant
$k_{\rm B}=1.0$). Time will be measured in units of $\sqrt{m\sigma_{\rm
AA}^2/48\epsilon_{\rm AA}}$, where $m$ is the mass of the particles.

The equations of motions have been integrated with the velocity form
of the Verlet algorithm, using a step size of 0.02. The number of A
and B particles were 800 and 200, respectively and the size of the box
was $(9.4)^3$. 

In the past the {\it equilibrium} dynamics of this system has been
determined in great detail~\cite{kob_lj,sastry98}. In particular it was shown
that at low temperatures the relaxation dynamics is described very well
by MCT with a critical temperature of $T_c=0.435$. To investigate the
aging dynamics we therefore equilibrated the system at the high initial
temperature $T_i=5.0$ and quenched it at time zero to a final temperature
$T_f\in \{0.1, 0.2, 0.3, 0.4, 0.435\}$. This quench was done by coupling
the particles every 50 time steps to a stochastic heat bath which was
kept on during the subsequent propagation of the system at low (kinetic)
temperature. In order to improve the statistics of the results we averaged
over 8-10 different realizations of the system.

\section{Results}

\begin{figure}[hbt]
\vspace*{80mm}
\includegraphics{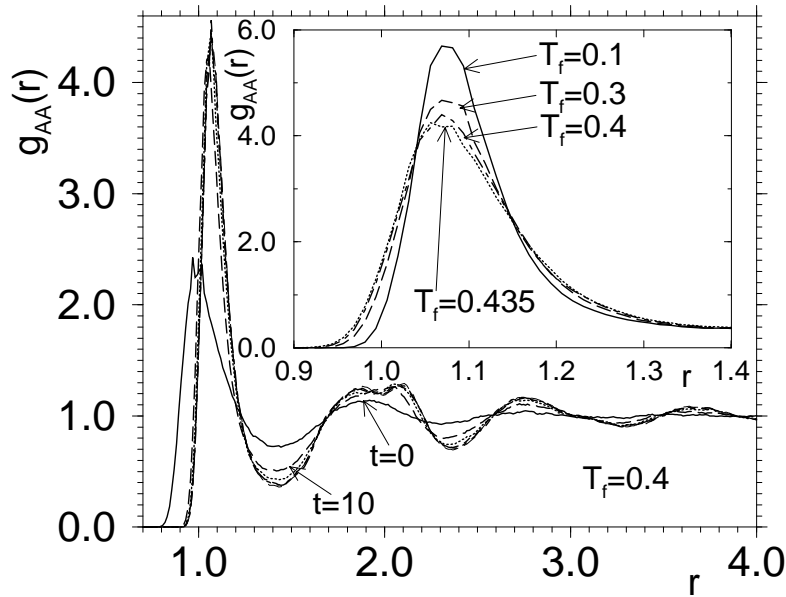}
\caption{
Main figure: radial distribution function between the A particles for
different times for $T_f=0.4$ . The times are $t=0$, i.e. before the
quench, $t=10, 100, 1000, 10000,$ and 63100 time units. Inset: The
same quantity at $t=63100$ for different values of $T_f$.
}
\label{fig1}
\end{figure}

Within the framework of the idealized version of MCT the aging process
is viewed as a slow approach of the system to the so-called ``critical
surface'' of MCT. This surface is a hyper-surface in the parameter space
of the coupling constants, which in the case of a structural glass
are given by the magnitude of $S(q)$, the static structure factor at
wave-vector $q$, and divides this space into a region in which the system
is liquid-like and one in which it is solid-like, i.e. a glass. (In
order to avoid some mathematical subtleties we consider
only a discrete and finite set of wave-vectors, thus the parameter space
is finite dimensional.) In order to check whether or not this surface
does indeed have any relevance for the aging dynamics of our system we
calculated the time dependence of $g_{\rm AA}(r)$, the radial distribution
function between two A particles, a quantity which is closely related
to the static structure factor. In Fig.~\ref{fig1} we show $g_{\rm AA}$
for different times $t$ after the quench (main figure). From this graph
we see that immediately after the quench the function changes rapidly
(compare the curves for $t=0$ and $t=10$) but soon afterwards shows only
a very weak time dependence and then can soon be considered as constant
within the accuracy of the data. That this limiting curve depends on the
final temperature $T_f$ of the quench is shown in the inset of the figure,
where we show $g_{\rm AA}(r)$ for other values of $T_f$. We see that
with decreasing $T_f$ the height of the main peak increases and that its
width decreases. The reason for this is that at such low temperatures
the particles vibrate in the cages formed by their neighbors and that the
size of this cage decreases with decreasing temperature. Note that this
dependence demonstrates that for different values of $T_f$ the system
populates states in different regions in configuration space. 

\begin{figure}[hbt]
\vspace*{80mm}
\includegraphics{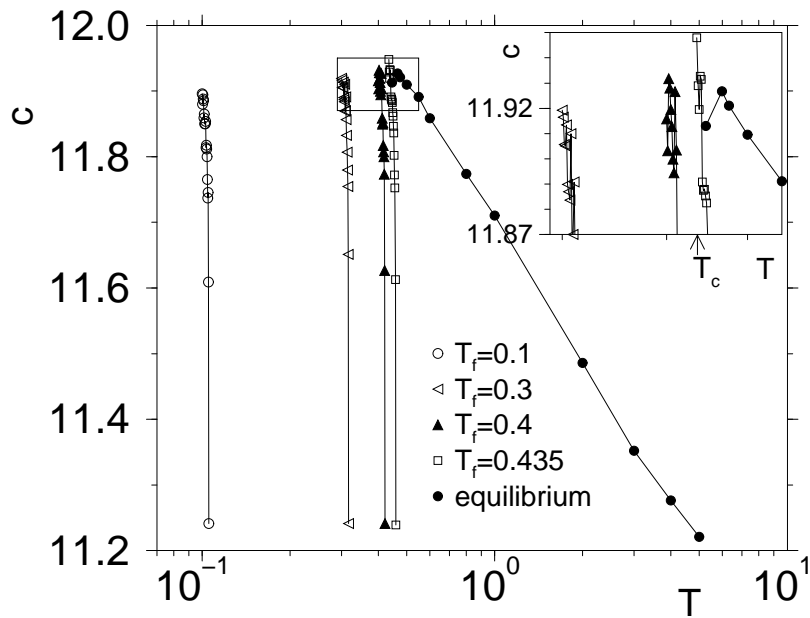}
\caption{
Main figure: time dependence of the area under the first peak in
$g_{\rm
AA}(r)$ for different values of $T_f$. The right-most curve is the
same
quantity for the {\it equilibrium} case. Inset: enlargement of the
equilibrium curve around the critical temperature of MCT.
}
\label{fig2}
\end{figure}

If the view put forward by MCT is correct these states should all
be close to the critical surface discussed above. In order to check
this prediction we calculated the area under the first peak in $g_{\rm
AA}(r)$.  This area is roughly proportional to the height of the main
peak in the static structure factor and previous calculations have
shown~\cite{bengtzelius84,nauroth97} that for simple systems such as the
present one this is the most relevant coupling parameter, i.e. the most
relevant direction in the parameter space of the coupling constants. In
Fig.~\ref{fig2} we show this area $c(t)$ for the different temperatures
$T_f$, i.e. $c(t)=4\pi\int_0^{r_c} r^2 g_{\rm AA}(t) dr$, where $r_c$
is the location of the minimum between the first and second peak in
$g_{\rm AA}(r)$. (Note that $c$ is nothing else than the average number
of A particles that surround an A particle.). The different symbols
correspond to times 0, 10, 40, 60, 100, 160, 250, 400, 630, 1000, 1580,
2510, 3980, 6310, 10000, 15850, 25120, 39810, and 63100 and are thus
space roughly equidistant on a logarithmic time axis. (In order to
expand the axis at low temperatures we chose a logarithmic temperature
scale.) From the main figure we recognize that at the beginning the area
does indeed depend strongly on time that, however, at intermediate and
long times $c(t)$ is essentially constant within the noise of the data
(see inset). We see that within the time scale covered the value of $c(t)$
changes only by about 6\%. That this small change is nevertheless very
significant is demonstrated by the right-most curve (filled circles)
in which the {\it equilibrium} value of $c$ at different temperatures is
shown~\cite{kob_lj,gleim98}.  From that curve we see that in the temperature
range $5.0\geq T \geq 0.446$ the area changes also only by about 6\%,
despite the fact that the dynamics of the system slows down by about
five orders of magnitude~\cite{kob_lj}.

The most relevant information from this graph is that the value of
$c(t)$ at long times seems to be almost independent of $T_f$, thus
giving evidence that at these times the system is indeed close to the
critical surface of MCT.  That this is indeed the critical surface can
be seen from the inset where we show an enlargement of the equilibrium
curve at low temperatures.  From previous investigations we know that at
these temperatures the (equilibrium) system is very close to its critical
temperature and that therefore the values of $c$ of the equilibrium curve
are very close to the critical ones and which are around 11.93. From the
inset we recognize that also the long time values of $c(t)$ are very close
to this number which is thus evidence that also the aging systems are,
at long times, quite close to the critical surface. A careful inspection
of the main figure reveals, however, that the curves for low values of
$T_f$ are a bit below this critical value, an observation which we will
discuss below.

\begin{figure}[hbt]
\vspace*{80mm}
\includegraphics{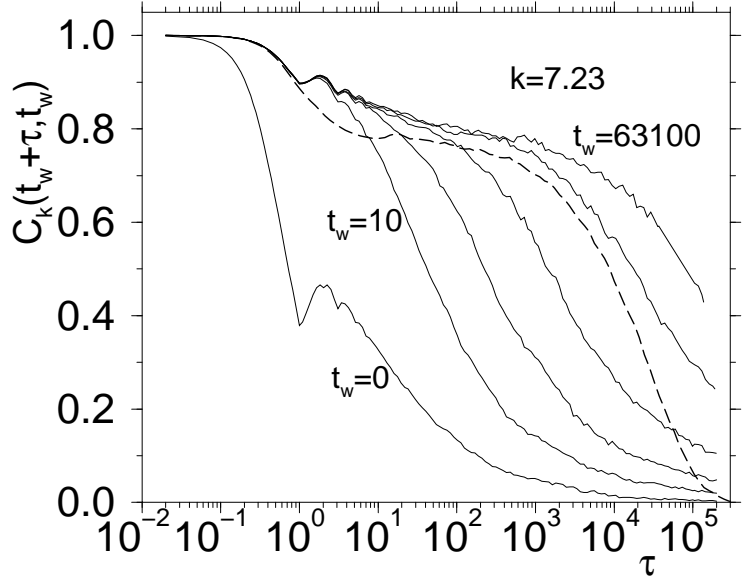}
\caption{
Time dependence of the generalization of the incoherent intermediate
scattering function to the out-of-equilibrium case. The different curves
correspond to different waiting times. $T_f=0.4$. The dashed curve is the
{\it equilibrium} curve at 0.446.
}
\label{fig3}
\end{figure}

From Figs.~\ref{fig1} and \ref{fig2} we see that during the aging
process the time dependence of the radial distribution function is rather
weak. This situation is typical for so-called ``one-time quantities'',
i.e. observables which {\it in equilibrium} are constant. (Below we will
discuss exceptions to this trend.) A much stronger time dependence is
found for the so-called ``two-time quantities'', i.e. the generalizations
of the equilibrium time correlation functions to the case of the out-of
equilibrium case. In equilibrium a time auto-correlation function of
an observable $y(t)$ depends only on the time difference, i.e. $\langle
y(t_w) y(t_w+\tau) \rangle=\langle y(\tau) y(0)\rangle$, where $\langle
. \rangle$ is the thermodynamic average. This equality no longer holds
for the out-of equilibrium case, since due to the generation of the
out-of equilibrium situation the time-translation invariance of the
system is lost. Therefore it is necessary to keep track of both times,
$t_w$, the time since the quench, and $\tau$, the time since the start
of the measurement. In the following we will study the case that the
observable is $\rho_s(k,t)$, the space-Fourier transform of the density
of a tagged at wave-vector $k$. This quantity is related to the positions
of the particles via
\begin{equation}
\delta \rho(k,t)=\frac{1}{N} \sum_{j=1}^N \exp[i{\bf k \cdot r}_j(t)]\quad .
\label{eq1}
\end{equation}
In equilibrium the resulting time correlation function is the so-called
incoherent intermediate scattering function $F_s(k,t)=\langle \rho(k,t)
\rho(-k,0) \rangle$ which can be measured in scattering experiments. For
the out-of equilibrium case we generalize this to
\begin{equation}
C_k(t_w+\tau,t_w)=\frac{1}{N}\sum_{j} \exp[i{\bf k}\cdot ({\bf
r}_j(t_w+\tau)-{\bf r}_j(\tau)]
\label{eq2}
\end{equation}
(Note that these last equations are trivially generalized to
multi-component systems.) In Fig.~\ref{fig3} we show the $\tau$ dependence
of $C_k(t_w+\tau,t_w)$ for the A particles for different waiting times
$t_w$. The value of the wave-vector is $k=7.23$, the location of the
main peak in the structure factor for the A-A correlation. (Other values
of $k$, as well as the curves for the B particles look qualitatively
similar.)  From this figure we recognize that this time correlation
function shows a very strong waiting time dependence, thus showing that
the investigation of aging effects is much easier when one considers
two-time quantities instead of one-time quantities (Figs.~\ref{fig1}
and \ref{fig2}). Note that this waiting time dependence is not found at
short times $\tau$, {\it if $t_w$ is large}, in that in this time regime
the different curves collapse onto a master curve. In this time regime
the particles are still inside the cage formed by their neighbors and
thus we conclude from the figure that this vibrational motion becomes
independent of the waiting time, if the latter is large. In Ref.~\cite{kob99}
evidence was given that the approach of the curves to the plateau is given
by a power-law, in agreement with the mean-field theories.

For larger times $\tau$ the particle starts to leave the mentioned cage
(the correlator starts to fall below the quasiplateau at intermediate
times) and from the figure we see that the time at which this happens
increases with increasing $t_w$ in that the time scale for the second
relaxation step increases rapidly with $t_w$. As it has been demonstrated
elsewhere~\cite{kob97}, this relaxation times scales roughly like
$t_w^{0.9}$, i.e. it shows a sub-aging behavior.

Also included in Fig.~\ref{fig3} is the {\it equilibrium} curve at
$T=0.446$ for the same wave-vector. Comparing this curve with the
out-of-equilibrium curves for large $t_w$ shows that the height of
the plateau is very similar. What is, however, very different is the
{\it second} relaxation process in that the equilibrium curve decays
much more rapidly, i.e has a larger slope, than the aging curves. A
detailed analysis shows that the former curve is approximated
well by a Kohlrausch-Williams-Watts function~\cite{kob_lj},
$\exp(-(\tau/\tau_0)^{\beta})$ whereas the latter curves are
power-laws, with an exponent that depends on $k$ but not on the waiting
time~\cite{kob99,kob99b}.

\begin{figure}[hbt]
\vspace*{80mm}
\includegraphics{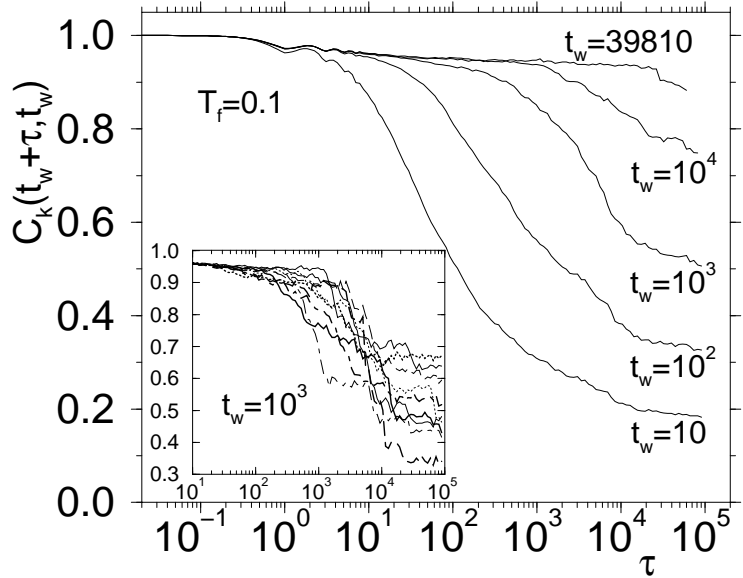}
\caption{
Time dependence of the generalization of the incoherent intermediate
scattering function to the out-of-equilibrium case for $T_f=0.1$
}
\label{fig4}
\end{figure}

The curves in Fig.~\ref{fig3} are for the final temperature
$T_f=0.4$, i.e. just about 10\% below the critical temperature of MCT
($T_c=0.435$). In order to see how $T_f$ affects the relaxation we
show in Fig.~\ref{fig4} the same type of correlation function as in
Fig.~\ref{fig3}, but this time for $T_f=0.1$.  Although the overall
behavior of these curves are similar to the ones for the higher $T_f$
some significant differences are found. First of all we see that the
height of the plateau, $f_c(k)$, is now quite a bit larger than the
one for $T_f=0.4$. This change can easily be understood by recalling
that for short times the motion of the particles is dominated by their
rattling inside the cage. To a first approximation this rattling can
be described by a superposition of harmonic oscillators and thus their
amplitude will be proportional to $T_f$. Thus we expect that $1-f_c(k)$ is
proportional to $T_f$, and an inspection of the curves in Figs.~\ref{fig3}
and \ref{fig4} shows that this is indeed the case. More noteworthy is
the observation that for large $t_w$ the curves seem to show a second
plateau at long times, which seems not to be present for $T_f=0.1$.

In order to understand the origin of this second plateau it is useful to
look at the individual runs for $T_f=0.1$, which are shown for $t_w=10^3$
in the inset of Fig.~\ref{fig4}. From this inset we see that the different
curves show a relatively sharp drop in the time range $2\times 10^2 \leq
\tau \leq 2\times 10^4$ and then is almost constant. Note that the time at
which this drop occurs depends on the realization. A careful analysis of
the configurations just before and after a sudden drop shows that this
fast relaxation is related to a very cooperative motion of about 10\%
of the particles~\cite{kob99}. Thus it seems that this mechanism is the
most effective way to release the stress that is in the configuration
due to the quench. This is in contrast to the behavior at higher values
of $T_f$ in that there the stress is smaller and thus the system can
remove it in a more gradual way, i.e. without the occurrence of the
``earthquakes'' that are seen at the lower temperatures.

In the remaining part of the paper we will discuss the aging of the system
from the point of view of the configuration space. For this we make use
of the concept of the ``inherent structure'' (IS), which was introduced
some time ago by Stillinger and Weber~\cite{inherent_structure} and can
be described as follows: Any point in configuration space can be used
as the starting point of a steepest descent procedure in the potential
energy of the system. The endpoint of this steepest descent is the IS
for the starting point. Thus in this way the configuration space can
be decomposed uniquely into the basins of attractions of the IS (apart
from some points of measure zero). By focusing on the IS we therefore
can study the evolution of the system during the aging process without
being disturbed by the vibrational part of the particle motion.

\begin{figure}[hbt]
\vspace*{80mm}
\includegraphics{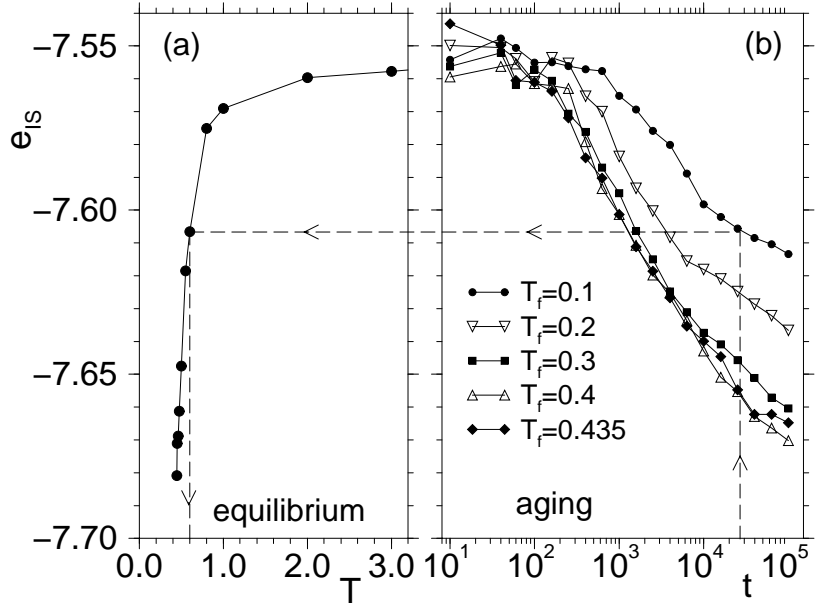}
\caption{
Time and time dependence of the energy of the IS. a): equilibrium, b)
out-of-equilibrium. Reproduced from Ref.~\protect\cite{kob99c}.
}
\label{fig5}
\end{figure}

In Fig.~\ref{fig5} we show the temperature dependence of $e_{\rm IS}$
the potential energy of the system in the IS {\it in equilibrium},
Fig.~\ref{fig5}a (see also Ref.~\cite{sastry98}). We recognize
that at high temperatures $e_{\rm IS}$ is basically constant and
starts to decrease quickly below $T\approx 1.0$ which shows that the
energy landscape, as characterized by the height of the local minima,
starts to change only when the system enters the supercooled regime. In
Fig.~\ref{fig5}b we show the {\it time} dependence of $e_{\rm IS}$ for
the different final temperatures investigated. We see that the curves for
small $T_f$ show three regimes. (Although the curves with higher $T_f$
show only two regimes we will argue below that also they should show a
third regime if one would be able to simulate for longer times.) The
first regime is observed at short times and in it $e_{\rm IS}(t)$ is
essentially independent of time.  After this time regime $e_{\rm IS}(t)$
enters the second regime during which the system is able to decreases
it energy. After a certain time $e_{\rm IS}$ crosses over to a weaker
time dependence and thus the system enters the third regime.

In order to understand these two last time regimes it is useful
to recall some results which have been obtained by analyzing
the instantaneous normal modes of supercooled liquids {\it in
equilibrium}~\cite{sciortino98}. Although these results have been obtained
for supercooled water it is likely that they can be transferred to the
present systems as well. What has been shown in Ref.~\cite{sciortino98}
is that the number of modes that lead the system to a {\it new} local
minimum decreases with decreasing temperatures and vanishes at the MCT
temperature $T_c$. Thus we can say that for temperatures above $T_c$
the typical configuration of the system has at least one unstable mode,
i.e. direction of motion, whereas for temperatures below $T_c$ the system
mainly sits in the vicinity of a local minimum, i.e. oscillates around a
metastable location. Thus at $T_c$ the system has a thermal energy which
is comparable with the difference in energy between $e_{\rm IS}(T_c)$ and
the lowest lying saddle point leading to a neighboring minimum. Hence, for
temperatures below $T_c$ the dynamics of the system becomes dominated by
activated processes.  Note that the fact that the temperature dependence
of the relaxation dynamics shows a strong deviation from an Arrhenius
temperature dependence~\cite{kob_lj,sciortino_water} implies that the height
of the {\it effective} barrier between two minima is not constant but
increases with decreasing temperature and, as explained above, at $T_c$
this barrier is on the order of $T_c$.

All this holds for the equilibrium case. For the out-of-equilibrium
case the situation is qualitatively similar but there is the important
difference that now the system has only the thermal energy $T_f$. The
three regimes seen in Fig.~\ref{fig5} can thus be explained as follows.
In the first regime the typical configurations of the system are
still quite similar to the ones found {\it in equilibrium} at high
temperatures, i.e. close to the initial temperature. In this part of
configuration space the effective barriers between adjacent minima are
relatively small but nevertheless noticeable. Since it takes the system
some time to find configurations with lower energy, $e_{\rm IS}$ does not
decrease. Only after some time the system manages to find configurations
which are energetically more favorable and hence $e_{\rm IS}$ starts
to decrease, i.e. in Fig.~\ref{fig5} $e_{\rm IS}(t)$ enters the second
regime. The time until such better configurations are found increases
with decreasing $T_f$, since a smaller kinetic energy makes it harder
for the system to cross the barriers.

During the aging process the system will lower its energy and start to
explore configurations which, {\it in equilibrium}, correspond to lower
and lower temperatures. After some time it will have reached that part
of configuration space in which the effective barriers to cross from
one minimum to the neighboring one have a height $k_BT_f$ and thus the
relaxation mechanism will change to an activated process. Since this
type of relaxation is less efficient than the one in which the system
still finds unstable modes, the rate by which the energy decreases is
decreasing. Thus the $e_{\rm IS}(t)$ curve shows a bend, which can be
seen in Fig.~\ref{fig5} when the system is entering the third regime.
Note that this bend in the $e_{\rm IS}(t)$ curves should occur at a
value of $e_{\rm IS}$ which increases with decreasing $T_f$ and this
is exactly what is seen in Fig.~\ref{fig5}. In particular we expect
from the reasoning above that if $T_f$ is very close to $T_c$ the third
regime should hardly be visible, and this expectation is indeed supported
by Fig.~\ref{fig5}.

\begin{figure}[hbt]
\vspace*{80mm}
\includegraphics{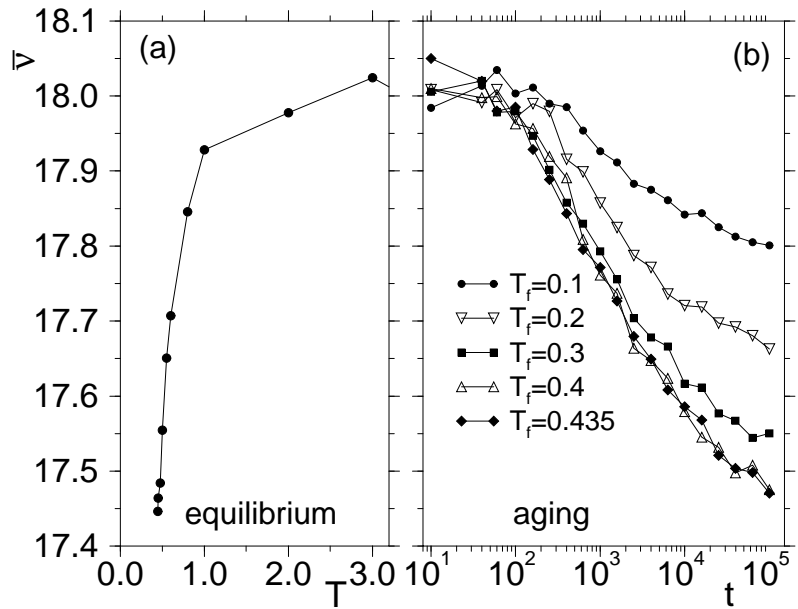}
\caption{
Time and time dependence of the first moment of the normal mode 
spectrum. a): equilibrium, b) out-of-equilibrium. Reproduced from
Ref.~\protect\cite{kob99c}.
}
\label{fig6}
\end{figure}

Since $e_{\rm IS}$ seems to be a quite sensitive quantity to locate the
place of the system in configuration space, and this is in contrast to
most other one-time quantities, we can use it to define an effective
temperature $T_e(t)$ during the aging process. For this we read off
the value of $e_{\rm IS}(t)$ of the aging system at a time $t$, and
define $T_e(t)$ to be that temperature $T$ at which the system {\it in
equilibrium} has the same value of $e_{\rm IS}$ (see Fig.~\ref{fig5}). In
order to check whether this definition of $T_e(t)$ has any physical
meaning it is necessary to show that from the knowledge of $T_e(t)$
it is possible to calculate other properties of the aging system. One
such property is e.g. the distribution of the frequencies of the normal
modes of the system. We have done this and found~\cite{kob99c} that the
value of $T_e(t)$ does indeed allow to calculate this distribution. In
the present paper we show, however, only $\bar{\nu}$, the first moment of
this distribution (see Fig.~\ref{fig6}) since it has a better statistical
accuracy than the distribution itself. In Fig.~\ref{fig6} we show
$\bar{\nu}$ for the equilibrium case as well as the aging case (left and
right panel, respectively). A comparison of these curves with the one
for $e_{\rm IS}$ in Fig.~\ref{fig5} shows immediately that the two sets
of curves are very similar. Hence it follows that $T_e(t)$ can indeed
be used to predict some of the properties of the aging system and thus
can indeed be considered as an effective temperature.

\section{Summary and Conclusions}
We have investigated the properties of a simple glass former after a
quench from a high temperature to a low temperature. From the radial
distribution function we have evidence that for long times after the
quench the system is very close to the critical surface of the MCT.
Note that, since this surface can be calculated from the structure
factor, it will {\it in the out-of-equilibrium situation} be a function
of $T_f$, since $S(q)$ does not only depend on the IS but also on the
thermal broadening of this configuration. This is the explanation why
during the aging the system is ``stuck'' in parts of configuration
space that depend on the value of $T_f$. At long times typical
configurations have the property that if their IS is thermally
broadened with a temperature $T_f$ the resulting structure factor is
very close to the critical surface of the MCT equations. Following the
conclusions of Ref.~\cite{sciortino98} one can rephrase this by saying
that for these configurations the typical barrier which leads to a
neighboring minimum is on the order of $T_f$.

Finally we mention that in a different place we have shown that the IS
can be used to determine the configurational entropy of the system at
low temperatures~\cite{sciortino99} which in turn allows to calculate
such interesting quantities as the Kauzman temperature.

Acknowledgments: We thank L. Cugliandolo, J. Kurchan, and A. Latz
for many helpful discussions. This work was supported by the Pole
Scientifique de Mod\'elisation Num\'erique at ENS-Lyon, and the Deutsche
Forschungsgemeinschaft through SFB 262.

\end{document}